\documentstyle[12pt]{article}
\begin{document}

\begin{center}

{\LARGE \bf Self-field $QED_{1+1}$ with massless matter fields:\\
Two-body problem}

\vspace{2 cm}

{\large \bf Fuad M. Saradzhev}

\vspace{1 cm}

{\it Institute of Physics, Academy of Sciences of Azerbaijan,\\
Huseyn Javid pr. 33, 370143 Baku, Azerbaijan}

\end{center}

\vspace{3 cm}

\rm

We consider two-body problem in the self-field $(1+1)$-
dimensional quantum electrodynamics on the circle.
We present two different formulations of the problem
which correspond to two different types of variational
principles and prove that both formulations lead to
the same spectrum of the two-body Hamiltonian with
massless matter fields. We give
the exact and complete solution of the relativistic
two-body equation in the massless case .

\vspace{2 cm}

PACS number(s) : 11.10.Kk , 12.20.Ds

\newpage

\begin{center}

\section{INTRODUCTION}

\end{center}

The relativistic two-body problem attracts attention
in both quantum mechanics of point particles and
field theory \cite{droz75}-\cite{wong93}. To formulate
the problem we usually assume that two or many body
systems are described by a composite field. This is
general in all formulations. What brings a difference
is variational principle.

We can rewrite the action of the two-body system
entirely in terms of the composite field and then
require the action to be stationary with respect to the
variations of this field only. This leads to a single
two-body equation \cite{oba86}. However, if we first vary
the action with respect to the individual fields, then
we come to a pair of coupled equations on the composite
field \cite{cra84,saz86}. These two different types of
variational principles produce therefore different types
of two-body equations.

In the present paper, we aim (i) to compare the two
formulations for the $(1+1)$-dimensional quantum
electrodynamics (QED) known as the Schwinger model
(SM) \cite{js62} and (ii) to solve exactly the two-body
problem for this model in the massless case. QED in lower
dimensions is interesting as a simpler model for discussion
of many body aspects of particle physics, for example,
spontaneous positron production by supercritical potentials
\cite{cal96}. Moreover, under certain conditions such lower
dimensions may be physically realizable in condensed matter
and statistical systems \cite{hoso97}. There is a discussion
of many-body problems in $(1+1)$-dimensions, however under
instantaneous phenomenological, e.g., $\delta$-functional
potentials \cite{glo87}.

We use the self-field version of QED \cite{barut83} which
is a first-quantized theory, so both matter and
electromagnetic fields are not quantized.
The electromagnetic field
has no separate local degrees of freedom and can be eliminated
between the coupled Maxwell-Dirac equations,
but then we must include
nonlinear self-field terms.
We consider two matter Dirac
fields coupled to a $U(1)$ gauge or electromagnetic field
and work on the
circle where the electromagnetic field has a global physical
degree of freedom.

Our paper is organized as follows. In Sec. 2 ,
for our two-body system  we present two alternative
formulations based on the variational principles
mentioned above and derive the corresponding
two-body equations in configuration space. We give
the Hamiltonian form of these equations in both
cases. The single two-body equation formulation is
one-time formulation. In contrast the formulation
with the composite field governed by two coupled
Dirac equations has two time coordinates and includes
the relative energy. In Sec. 3, we find the eigenfunctions
and the spectrum of the two-body Hamiltonian in the
single two-body equation formulation with massless
matter fields. In \cite{barsa94} we solved this problem
with the self-potentials neglected. Now we treat
the case of the massless matter fields completely.
We take into account in the two-body equation the
self-potentials responsible for the radiative corrections
and solve it exactly, i.e., get the exact and complete
solution of the two-body problem. In Sec. 4, we consider
the eigenvalue problem for the two-body Hamiltonian
in the pair of Dirac equations formulation as well. We
prove that both formulations lead to the same spectrum
and are therefore equivalent to each other. Sec. 5
contains our conclusions.

\newcommand{\ren}{\renewcommand{\theequation}{2.\arabic{equation}}}
\newcommand{\new}[1]{\renewcommand{\theequation}{2.\arabic{equation}#1}}
\newcommand{\add}{\addtocounter{equation}{-1}}

\begin{center}

\section{TWO-BODY SYSTEM}

\end{center}

The action of the system is
\ren
\begin{equation}
W[{\psi},A] = \int_{-\infty}^{\infty} dt \int_{0}^{\rm L} dx
\sum_{k=1}^{2} \{ [ \overline{\psi}_k {\gamma}^{\mu}
(i {\partial}_{\mu} -e_k A_{\mu}) {\psi}_k  - m_k \overline{\psi}_k
{\psi}_k] - \frac{1}{4} F_{\mu \nu} F^{\mu \nu} \} ,
\label{eq: dvaodin}
\end{equation}
where $(\mu,\nu = \overline{0,1})$, ${\gamma}^0=-i{\sigma}_2$,
${\gamma}^0{\gamma}^1={\gamma}^5={\sigma}_3$,
${\sigma}_i$ $(i=\overline{1,3})$ are Pauli matrices. The
fields ${\psi}_k$ are $2$-component Dirac spinors, and
$\bar{\psi}_k ={\psi}_k^{\star}{\gamma}^0$.

We suppose that space is a circle of length ${\rm L}$,
$0 \leq x <{\rm L}$ , and impose the following boundary
conditions for the fields
\begin{eqnarray*}
A_{\mu}({\rm L},t) & = & A_{\mu}(0,t), \\
{\psi}_k({\rm L},t) & = & e^{i2{\pi}{\kappa}_k} {\psi}_k(0,t),
\hspace{1 cm} k=1,2,
\end{eqnarray*}
${\kappa}_1, {\kappa}_2$  being arbitrary numbers.
The charges $e_1$ , $e_2$ are not arbitrary
on the circle, one of the charges must be a multiple of
another \cite{sara97}.

We work in the Coulomb gauge
\[
A_1(x,t) =b(t),
\]
where
\[
b(t) \equiv \frac{1}{\rm L} \int_{0}^{\rm L} dx
A_1(x,t)
\]
is the electromagnetic  field global degree of freedom.

The electromagnetic field equations deduces from the action
(\ref{eq: dvaodin}) are
\ren
\begin{equation}
{\partial}_{\nu} F^{\nu \mu} =  J^{\mu},
\label{eq: dvadva}
\end{equation}
where the total matter current
\[
J^{\mu} =\sum_{k=1}^2 e_k \bar{\psi}_k {\gamma}^{\mu} {\psi}_k
\]
is conserved , ${\partial}_{\mu} J^{\mu} =0$.

If we solve the electromagnetic field equations, express $A_{\mu}$
in terms of $J^{\mu}$ and insert the expressions obtained into
(\ref{eq: dvaodin}), then we get the action written in terms of the
matter fields
\[
W[{\psi},A] = \int_{- \infty}^{\infty} dt \int_{0}^{\rm L} dx
\sum_{k=1}^{2}  \overline{\psi}_k ({\gamma}^{\mu} i
{\partial}_{\mu} - m_k) {\psi}_k + \frac{1}{2}
\int_{- \infty}^{\infty} dt \int_{0}^{\rm L} dx \int_{0}^{\rm L}
dy J^0(x,t) D(x,y|{\rm L}) J^0(y,t)
\]
\ren
\begin{equation}
- \frac{1}{2} \int_{- \infty}^{\infty} dt \int_{0}^{\rm L} dx
J^1(x,t) b(t).
\label{eq: dvatri}
\end{equation}
The last term represents the interaction of the matter currents
with the global electromagnetic field degree of freedom $b$,
while the middle term is a sum of current-current interactions
containing both the mutual and self-interaction terms.

The Green's function in (\ref{eq: dvatri}) is
\[
D(x,y|{\rm L}) \equiv \frac{1}{2} |x-y| + \frac{xy}{\rm L}.
\]

\begin{center}

\subsection{First formulation: single two-body equation}

\end{center}

Following the relativistic configuration space formalism
\cite{oba86}, we define the composite field
\[
{\Phi}(x_1,t|x_2,t) \equiv {\psi}_1(x_1,t) \otimes   {\psi}_2(x_2,t),
\]
which is $4$-component spinor field. The configuration space
$(x_1,x_2)$  is a torus with the circle length
$(0 \leq x_1 <{\rm L}$ , $0 \leq x_2 <{\rm L})$ .

We can rewrite our action (\ref{eq: dvatri}) entirely in terms
of the composite field ${\Phi}$. In order to do this we multiply
the kinetic energy terms with the normalization factors
\ren
\begin{equation}
\int_{0}^{\rm L} dx {\psi}^{\star}_k(x,t) {\psi}_k(x,t) =1,
\hspace{5 mm} k=1 \hspace{3 mm} {\rm or} \hspace{3 mm} 2.
\label{eq: dvacet}
\end{equation}
We have to do this twice on the self-interaction terms.
The resultant action is
\[
{\rm W}[\Phi, A] = \int_{- \infty}^{\infty} dt \int_{0}^{\rm L} dx_1
\int_{0}^{\rm L} dx_2 \bar{\Phi}(x_1,t|x_2,t)
\{ ({\gamma}^{\mu} p_{(1),\mu} -m_1) \otimes {\gamma}^0
+{\gamma}^0 \otimes ({\gamma}^{\mu} p_{(2),\mu} -m_2)
\]
\[
+ \frac{1}{2} ({\gamma}^0 \otimes {\gamma}^0) (e_1 {\phi}^{\rm self}_{(1)}
+ e_2 {\phi}^{\rm self}_{(2)}) - \frac{1}{2} e_1 b ({\gamma}^1
\otimes {\gamma}^0 ) - \frac{1}{2} e_2 b ({\gamma}^0 \otimes
{\gamma}^1 )
\]
\ren
\begin{equation}
+ e_1e_2 ({\gamma}^0 \otimes {\gamma}^0)
D(x_1,x_2|{\rm L}) \} {\Phi}(x_1,t|x_2,t) ,
\label{eq: dvapet}
\end{equation}
where
\[
p_{(i),\mu} \equiv  i \frac{\partial}{\partial x_{i}^{\mu}} ,
\]
and
\[
{\phi}_{(1)}^{\rm self}(x_1 ,t)
= e_1 \int_0^{\rm L} dy \int_0^{\rm L} dz D(x_1,z|{\rm L})
\bar{\Phi}(z,t|y,t) ({\gamma}^0 \otimes {\gamma}^0)
{\Phi}(z,t|y,t),
\]
\[
{\phi}_{(2)}^{\rm self}(x_2 ,t)
= e_2 \int_0^{\rm L} dy \int_0^{\rm L} dz D(x_2,y|{\rm L})
\bar{\Phi}(z,t|y,t) ({\gamma}^0 \otimes {\gamma}^0)
{\Phi}(z,t|y,t),
\]
the self-potentials ${\phi}_{(k)}^{\rm self}$ being
non-linear integral expressions.
The spin matrices are written here in the form of tensor
products $\otimes$, the first factor always referring to the
spin space of particle 1, the second to particle 2.

Let us note that the last term in (\ref{eq: dvapet})
can also be put into the
self-potentials ${\phi}_{(k)}^{\rm self}$, one half
for each particle; the total potentials then take the form
\[
{\phi}_{(1)}^{\rm self} \rightarrow {\phi}^{\rm self},
\]
\[
{\phi}_{(2)}^{\rm self} \rightarrow {\phi}^{\rm self},
\]
where
\ren
\begin{equation}
{\phi}^{\rm self}(x,t) \equiv \int_{0}^{\rm L} dy
\int_{0}^{\rm L} dz (e_1 D(x,z|{\rm L}) + e_2 D(x,y|{\rm L}) )
\overline{\Phi}(z,t|y,t) ({\gamma}^0 \otimes {\gamma}^0)
{\Phi}(z,t|y.t).
\label{eq: dvashest}
\end{equation}

We must now specify a variational principle for the matter
fields. We could vary the action with respect to individual
fields ${\psi}_1$ and ${\psi}_2$ separately. This results in
non-linear coupled equations for these fields (see below).
Instead, we require the action (\ref{eq: dvapet}) to be
stationary with respect
to the total composite field only. This is a weaker condition
which leads to the
following two-body equation
\[
\{ ({\gamma}^{\mu} p_{(1),\mu} - m_1) \otimes {\gamma}^0 +
{\gamma}^0 \otimes ({\gamma}^{\mu} p_{(2),\mu} -m_2)
+ ({\gamma}^0 \otimes {\gamma}^0) (e_1 {\phi}_{(1)}^{\rm self}
+ e_2 {\phi}_{(2)}^{\rm self} )
\]
\ren
\begin{equation}
- \frac{1}{2} e_1 b
({\gamma}^1 \otimes {\gamma}^0) - \frac{1}{2} e_2 b
({\gamma}^0 \otimes {\gamma}^1)
+ e_1e_2 ({\gamma}^0 \otimes {\gamma}^0) D(x_1,x_2|{\rm L}) \}
{\Phi}(x_1,t|x_2,t) =0.
\label{eq: dvasem}
\end{equation}

If we define the generalized (kinetic) momenta as
\[
{\pi}_{(i),\mu} \equiv p_{(i),\mu} + e_i A_{(i),\mu}^{\rm self}
\]
with
\[
A_{(1),0}^{\rm self} \equiv {\phi}_{(1)}^{\rm self}  ,
\hspace{1 cm}
A_{(2),0}^{\rm self} \equiv {\phi}_{(2)}^{\rm self} ,
\]
\[
A_{(1),1}^{\rm self} = A_{(2),1}^{\rm self} = - \frac{1}{2} b,
\]
then the two-body equation takes the compact form
\ren
\begin{equation}
\{ ({\gamma}^{\mu} {\pi}_{(1),\mu} - m_1 ) \otimes {\gamma}^0
+ {\gamma}^0 ( {\gamma}^{\mu} {\pi}_{(2),\mu} - m_2) +
e_1e_2 ({\gamma}^0 \otimes {\gamma}^0) D(x_1,x_2|{\rm L}) \}
{\Phi}(x_1,t|x_2,t) =0.
\label{eq: dvavosem}
\end{equation}

In the center of mass and relative coordinates
\begin{eqnarray*}
{\Pi} = {\pi}_{(1)} + {\pi}_{(2)} & , & {\pi} = {\pi}_{(1)} - {\pi}_{(2)}, \\
P=p_{(1)} + p_{(2)} & , & p = p_{(1)} - p_{(2)}, \\
x_{+} = x_1 + x_2 & , & x_{-} = x_1 - x_2 , \\
\end{eqnarray*}
the configuration space $(x_{-},x_{+})$
is again a torus, but with the circle length
$2{\rm L}$ $( -{\rm L} \leq x_{-} < {\rm L}$,
$0 \leq x_{+} < 2{\rm L})$ , while the function
$D(x_1,x_2|{\rm L})$ becomes a sum of center of mass
and relative parts depending only on $x_{-}$ and $x_{+}$ ,
respectively,
\begin{eqnarray*}
D(x_1,x_2|{\rm L}) & = & D_{-}(x_{-}|{\rm L}) +  D_{+}(x_{+}|{\rm L}),\\
D_{-}(x_{-}|{\rm L}) & = & \frac{1}{2} |x_{-}| -
\frac{1}{4{\rm L}} x_{-}^2 ,\\
D_{+}(x_{+}|{\rm L}) & = & \frac{1}{4{\rm L}} x_{+}^2 .\\
\end{eqnarray*}

Eq.(\ref{eq: dvavosem}) , without the self-field terms, becomes
\ren
\begin{equation}
\{ {\Gamma}^{\mu} P_{\mu} + k^{\mu} p_{\mu} + e_1e_2
({\gamma}^0 \otimes {\gamma}^0) D - m_1 (I \otimes {\gamma}^0)
- m_2 ({\gamma}^0 \otimes I) \} {\Phi}(x_{-},t|x_{+},t)=0,
\label{eq: dvadevet}
\end{equation}
where we have introduced
\[
{\Gamma}^{\mu} \equiv \frac{1}{2} ({\gamma}^{\mu} \otimes
{\gamma}^0 + {\gamma}^0 \otimes {\gamma}^{\mu} )
\]
\[
k^{\mu} \equiv \frac{1}{2} ({\gamma}^{\mu} \otimes {\gamma}^0
- {\gamma}^0 \otimes {\gamma}^{\mu} ),
\]
and $I$ is identity matrix.
Since $k^0$ vanishes, the zero component of $p_{\mu}$ , i.e.,
the relative energy $p_0$ drops out of the two-body equation
automatically. Thus we have only one time variable conjugate
to the center of mass energy $P_0$ , one degree of freedom
for the center of mass momentum $P^1$ and one degree of
freedom for the relative momentum $p^1$. By multiplying
(\ref{eq: dvadevet})
by ${\Gamma}_0^{-1}$ we obtain the Hamiltonian form
of the two-body equation
\ren
\begin{equation}
P_0 {\Phi} = \{ {\alpha}_{+} P^1 + {\alpha}_{-}
p^1 - e_1e_2 D + {\beta}_1 m_1 +
{\beta}_2 m_2  \} {\Phi},
\label{eq: dvadeset}
\end{equation}
with
\[
{\alpha}_{\pm} \equiv \frac{1}{2} ({\alpha}_1 {\pm} {\alpha}_2),
\hspace{5 mm}
{\alpha}_1 \equiv {\gamma}^5 \otimes I,
\hspace{5 mm}
{\alpha}_2 \equiv I \otimes {\gamma}^5,
\]
\[
{\beta}_1 \equiv {\gamma}^0 \otimes I ,
\hspace{5 mm}
{\beta}_2 \equiv I \otimes {\gamma}^0 ,
\]
and the relative and center of mass terms in the Hamiltonian
$P_0$ being additive,
\[
P_0 = H_{\rm c.m.} + H_{\rm rel},
\]
\[
H_{\rm c.m.} \equiv {\alpha}_{+} P^1 - e_1e_2 D_{+},
\]
\[
H_{\rm rel} \equiv {\alpha}_{-} p^1
- e_1e_2D_{-} + {\beta}_1 m_1 + {\beta}_2 m_2.
\]
Eq.(\ref{eq: dvadeset}) has the form of a generalized
Dirac equation, now a $4$-component wave equation.

The commutation relations for the matrices ${\alpha}_{\pm}$,
${\beta}_1$, ${\beta}_2$ are
\[
{\alpha}_{+} {\alpha}_{-} = {\alpha}_{-} {\alpha}_{+} =0,
\hspace{5 mm}
{\beta}_1 {\beta}_2 = {\beta}_2 {\beta}_1 =
{\gamma}^0 \otimes {\gamma}^0,
\]
\[
{\alpha}_{\pm} {\beta}_1 + {\beta}_1 {\alpha}_{\pm}
= \pm {\gamma}^0 \otimes {\gamma}^5,
\hspace{5 mm}
{\alpha}_{\pm} {\beta}_2 + {\beta}_2 {\alpha}_{\pm}
= {\gamma}^5 \otimes {\gamma}^0,
\]
and
\[
{\alpha}_{\pm}^2 = \frac{1}{2} (I \otimes I \pm
{\gamma}^5 \otimes {\gamma}^5),
\hspace{5 mm}
{\beta}_1^2 = {\beta}_2^2 = I \otimes I.
\]

With the self-potential terms included, the Hamiltonian form
of the two-body equation becomes
\ren
\begin{equation}
P_0 {\Phi}_0 = \{ {\alpha}_{+} {\Pi}^1 +
{\alpha}_{-} {\pi}^1 - {\phi} - e_1 {\phi}_{(1)}^{\rm self}
- e_2 {\phi}_{(2)}^{\rm self} + {\beta}_1 m_1 +
{\beta}_2 m_2 \} {\Phi},
\label{eq: dvaodinodin}
\end{equation}
where
\[
{\phi} = {\phi}_{+} + {\phi}_{-},
\]
\[
{\phi}_{\pm} \equiv e_1e_2 D_{\pm}.
\]
The self-potentials break in general the above mentioned
additivity of the center of mass and relative parts of
$P_0$.

\begin{center}

\subsection{Second formulation: pair of Dirac equations}

\end{center}

Let us use now a different variational principle and
vary the action (\ref{eq: dvatri}) with respect to each field
${\psi}_k$ separately. In this way we come to a pair of coupled
nonlinear equations
\new{a}
\begin{equation}
({\gamma}^{\mu} i {\partial}_{\mu} - m_1){\psi}_1(x,t) -
\frac{1}{2} e_1 b(t) {\gamma}^1 {\psi}_1(x,t) +
e_1 \int_{0}^{\rm L} dy D(x,y|{\rm L}) J^0(y,t)
{\gamma}^0 {\psi}_1(x,t) =0,
\end{equation}
\add
\new{b}
\begin{equation}
({\gamma}^{\mu} i{\partial}_{\mu} -m_2) {\psi}_2(x,t)
- \frac{1}{2} e_2 b(t) {\gamma}^1 {\psi}_2(x,t)
+ e_2 \int_{0}^{\rm L} dy D(x,y|{\rm L}) J^0(y,t)
{\gamma}^0 {\psi}_2(x,t) =0.
\end{equation}

To describe our two-body system we define the composite field
\[
{\Phi}(x_1,t_1|x_2,t_2) = {\psi}_1(x_1,t_1) \otimes
{\psi}_2(x_2,t_2)
\]
composed of the individual matter fields at different
times.
Multiplying Eq.(2.12a) taken at $(x,t)=(x_1,t_1)$ by
${\gamma}^0{\psi}_2(x_2,t_2)$ and Eq.(2.12b) taken at
$(x,t)=(x_2,t_2)$ by ${\gamma}^0{\psi}_1(x_1,t_1)$
as well as the nonlinear self-field terms in both
equations by the normalization factors leads to
\[
G_1 {\Phi}(x_1,t_1|x_2,t_2) \equiv \{ ({\gamma}^{\mu}
p_{(1),\mu} -m_1) \otimes {\gamma}^0 - \frac{1}{2}
e_1 b(t_1) ({\gamma}^1 \otimes {\gamma}^0)
\]
\new{a}
\begin{equation}
+ ({\gamma}^0 \otimes {\gamma}^0) e_1 {\phi}^{\rm self}(1) \}
{\Phi}(x_1,t_1|x_2,t_2) =0,
\end{equation}
\[
G_2 {\Phi}(x_1,t_1|x_2,t_2) \equiv \{ {\gamma}^0 \otimes
({\gamma}^{\mu} p_{(2),\mu} -m_2) - \frac{1}{2}e_2 b(t_2)
({\gamma}^0 \otimes {\gamma}^1)
\]
\add
\new{b}
\begin{equation}
+ ({\gamma}^0 \otimes
{\gamma}^0) e_2 {\phi}^{\rm self}(2) \}
{\Phi}(x_1,t_1|x_2,t_2)=0,
\end{equation}
where the self-potential is
\[
{\phi}^{\rm self}(x|t_1,t_2) \equiv \int_0^{\rm L} dy
\int_0^{\rm L} dz D(x,y|{\rm L}) \{ e_1 \overline{\Phi}
(y,t_1|z,t_2) ({\gamma}^0 \otimes {\gamma}^0) {\Phi}(y,t_1|
z,t_2)
\]
\[
+ e_2 \overline{\Phi}(z,t_2|y,t_1) ({\gamma}^0
\otimes {\gamma}^0) {\Phi}(z,t_2|y,t_1) \},
\]
and
\[
{\phi}^{\rm self}(1) \equiv {\phi}^{\rm self}(x_1|t_1,t_2),
\hspace{5 mm}
{\phi}^{\rm self}(2) \equiv {\phi}^{\rm self}(x_2|t_2,t_1),
\]
i.e., we have a pair of Dirac equations on ${\Phi}$ instead
of a single one in the first formulation.
For $t_1=t_2 \equiv t$, ${\phi}^{\rm self}(x|t,t)$ coincides
with the self-potential ${\phi}^{\rm self}(x,t)$ given
by (\ref{eq: dvashest}).

The compatibility condition for the two equations is
\ren
\begin{equation}
[G_1 , G_2]_{-} {\Phi} =0.
\label{eq: dvaodincet}
\end{equation}
It can be checked that this condition reduces to
\begin{equation}
e_1 \frac{{\partial}{\phi}^{\rm self}(1)}{{\partial}t_2}
= e_2 \frac{{\partial}{\phi}^{\rm self}(2)}{{\partial}t_1},
\label{eq: dvaodinpet}
\end{equation}
i.e. requires a specific time dependence of the self-potentials.

Taking the sum and the difference of Eqs.(2.13a-b) we get
\[
\{ ({\gamma}^{\mu} p_{(1),\mu} - m_1) \otimes {\gamma}^0
+ {\gamma}^0 \otimes ({\gamma}^{\mu} p_{(2),\mu} - m_2) +
({\gamma}^0 \otimes {\gamma}^0) (e_1 {\phi}^{\rm self}(1)
+ e_2 {\phi}^{\rm self}(2) )
\]
\ren
\new{a}
\begin{equation}
- \frac{1}{2} e_1 b(t_1) ({\gamma}^1 \otimes {\gamma}^0 )
- \frac{1}{2} e_2 b(t_2) ({\gamma}^0 \otimes {\gamma}^1 ) \}
{\Phi}(x_1,t_1|x_2,t_2) =0,
\end{equation}
and
\[
\{({\gamma}^{\mu} p_{(1),\mu} - m_1) \otimes {\gamma}^0
- {\gamma}^0 \otimes ({\gamma}^{\mu} p_{(2),\mu} - m_2 ) +
({\gamma}^0 \otimes {\gamma}^0) (e_1 {\phi}^{\rm self}(1)
- e_2 {\phi}^{\rm self}(2) )
\]
\ren
\add
\new{b}
\begin{equation}
- \frac{1}{2} e_1 b(t_1) ({\gamma}^1 \otimes {\gamma}^0)
+ \frac{1}{2} e_2 b(t_2) ({\gamma}^0 \otimes {\gamma}^1) \}
{\Phi}(x_1,t_1|x_2,t_2) =0.
\end{equation}
The first equation is in fact the two-body equation derived
earlier with the Coulomb potential included into the
self-potentials, the only difference being in the number of time
variables, while (2.16b) is a new equation on ${\Phi}$.

To make clear the nature of the new equation, we use again
the center of mass and relative coordinates. Acting along
similar lines as above, we obtain the Hamiltonian form of the
equations on ${\Phi}$ :
\new{a}
\begin{equation}
P_0 {\Phi} = \{ {\alpha}_{+} {\Pi}^1 + {\alpha}_{-} {\pi}^1
-e_1 {\phi}^{\rm self}(1) - e_2 {\phi}^{\rm self}(2) +
{\beta}_1 m_1 + {\beta}_2 m_2 \} {\Phi},
\end{equation}
\add
\new{b}
\begin{equation}
p_0 {\Phi} = \{ {\alpha}_{+} {\pi}^1 + {\alpha}_{-} {\Pi}^1
-e_1 {\phi}^{\rm self}(1) + e_2 {\phi}^{\rm self}(2)
+ {\beta}_1 m_1 - {\beta}_2 m_2 \} {\Phi} .
\end{equation}
In addition to the two-body equation we have therefore
an equation which includes the relative energy $p_0$. While
the center of mass energy plays the role of the "Hamiltonian"
of the two-body system, the relative energy (or its conjugate
variable, the relative time) is an unphysical variable and
must be eliminated to avoid possible unphysical effects, for
example, relative energy excitations in the spectrum.

In the spectrum problem we can simply put
\ren
\begin{equation}
p_0 {\Phi} =0,
\label{eq: dvaodinvosem}
\end{equation}
i.e., assume that ${\Phi}$ does not depend on the relative time
$\tau = t_1-t_2$. We could also start from the beginning with
the field ${\Phi}$ composed of the individual matter fields
taken at the same time $t_1=t_2=t$. Then the compatibility
condition of the two Dirac equations would be
\begin{equation}
\frac{{\partial}}{{\partial} t} (e_1 {\phi}^{\rm self}(x_1,t)
- e_2 {\phi}^{\rm self}(x_2,t) )=0.
\label{eq: dvaodindevet}
\end{equation}
We shall continue our discussion of the pair of Dirac
equations formulation in Sec. 4.

\renewcommand{\ren}{\renewcommand{\theequation}{3.\arabic{equation}}}
\renewcommand{\add}{\addtocounter{equation}{-1}}
\renewcommand{\new}[1]{\renewcommand{\theequation}{3.\arabic{equation}#1}}
\newcommand{\set}{\setcounter{equation}{0}}

\begin{center}

\section{MASSLESS CASE}

\end{center}

\set

There are three types of interactions in the first quantized
two-body Hamiltonian $P_0$, namely, interaction described
by the self-potentials ,
interaction between the matter fields and global electromagnetic
field degree of freedom and
the Coulomb interaction .
All these interactions influence
the spectrum, the self-potentials being responsible for
radiative processes.

Let us find the eigenfunctions and the spectrum of $P_0$
in the single two-body equation formulation.
The consideration below is at fixed time $t=0$.
The equation for the eigenfunctions is
\ren
\begin{equation}
({\alpha}_{+} {\Pi}^1 + {\alpha}_{-} {\pi}^1
+ {\beta}_1 m_1 + {\beta}_2 m_2 ) {\Phi} =
(E+V) {\Phi},
\label{eq: triodin}
\end{equation}
where
\[
{\Pi}^1 = 2i \frac{\partial}{\partial x_{+}} -
\frac{1}{2} (e_1+e_2)b,
\]
\[
{\pi}^1 = 2i \frac{\partial}{\partial x_{-}} -
\frac{1}{2} (e_1-e_2)b,
\]
and
\[
V(x_{-},x_{+}) = {\phi} + e_1{\phi}_{(1)}^{\rm self} +
e_2 {\phi}_{(2)}^{\rm self}.
\]

If we denote the components of ${\Phi}$ as
\begin{eqnarray*}
{\Phi}^{11} \equiv {\eta}_1 & , & {\Phi}^{12} \equiv {\eta}_2,\\
{\Phi}^{21} \equiv {\eta}_3 & , & {\Phi}^{22} \equiv {\eta}_4,\\
\end{eqnarray*}
then (\ref{eq: triodin}) reduces to the system of four equations
\[
2i \frac{\partial}{\partial x_{+}} {\eta}_1 -
(V+E + \frac{1}{2}(e_1+e_2)b ) {\eta}_1
=  -m_1 {\eta}_3 - m_2 {\eta}_2,
\]
\ren
\begin{equation}
2i \frac{\partial}{\partial x_{+}} {\eta}_4 +
(V +E -\frac{1}{2}(e_1+e_2)b ) {\eta}_4
=  m_1 {\eta}_2 + m_2 {\eta}_3,
\label{eq: tridva}
\end{equation}
\[
2i \frac{\partial}{\partial x_{-}} {\eta}_2 -
(V+E + \frac{1}{2}(e_1-e_2)b ) {\eta}_2
=  -m_1 {\eta}_4 - m_2 {\eta}_1,
\]
\ren
\begin{equation}
2i \frac{\partial}{\partial x_{-}} {\eta}_3 +
(V+E - \frac{1}{2}(e_1-e_2)b ) {\eta}_3
=  m_1 {\eta}_1 + m_2 {\eta}_4.
\label{eq: tritri}
\end{equation}
The global electromagnetic field degree of freedom shows itself in
all four equations. For $e_1=-e_2$ , $b$
drops out of the first pair of the equations, and for
$e_1=e_2$  of the second one.

We see from these equations that
\new{a}
\begin{equation}
{\eta}_1^{\star}(E,-e_1,-e_2) =  {\eta}_4(E,e_1,e_2),
\end{equation}
\add
\new{b}
\begin{equation}
{\eta}_2^{\star}(E,-e_1,-e_2) =  {\eta}_3(E,e_1,e_2),
\end{equation}
so only half of all solutions correspond to physical
particles.

The conditions (3.4a-b) are modified in the case
of the massless matter fields and vanishing total potential
$V$,
\begin{eqnarray*}
{\eta}_1(-E,e_1,e_2) & = & {\eta}_4 (E,e_1,e_2),\\
{\eta}_2(-E,e_1,e_2) & = & {\eta}_3 (E,e_1,e_2),\\
\end{eqnarray*}
i.e., the negative energy solutions of ${\eta}_1$ and
${\eta}_2$ coincide correspondingly with the positive
energy solutions of ${\eta}_4$ and ${\eta}_3$.
Again only half of all solutions correspond to physical
particles.

The boundary and normalization conditions for ${\eta}_{i}$
$(i=\overline{1,4})$ deduced from the ones for the
individual matter fields are
\begin{eqnarray*}
{\eta}_i({\rm L}|{\rm L}) & = & \exp \{ i2{\pi}{\kappa}_1^{(i)} \}
{\eta}_i(0|0) ,\\
{\eta}_i(-{\rm L}|{\rm L}) & = & \exp\{ i2{\pi}{\kappa}_2^{(i)} \}
{\eta}_i(0|0) ,\\
{\eta}_i(0|2{\rm L}) & = & \exp \{ i2{\pi} ({\kappa}_1^{(i)} +
{\kappa}_2^{(i)}) \} {\eta}_i(0|0) ,\\
\end{eqnarray*}
and
\[
\int_{-{\rm L}}^{\rm L} dx_{-} \int_{0}^{2{\rm L}} dx_{+}
{\eta}_i^{\star}(x_{-}|x_{+}) {\eta}_i(x_{-}|x_{+}) =1 ,
\]
respectively (no summation over i).

For the massless matter fields, $m_1=m_2=0$ and ${\eta}_{i}$
decouple from each other in Eqs.(\ref{eq: tridva}) - (\ref{eq: tritri})
which are therefore simplified. In what follows
we consider in detail the two-body problem for the massless
matter fields.

\begin{center}

\subsection{The case ${\phi}_{(1)}^{\rm self} =
{\phi}_{(2)}^{\rm self}=0$}

\end{center}

In \cite{barsa94} we put the self-potentials
${\phi}_{(k)}^{\rm self}$ equal to zero
and solved Eqs.(\ref{eq: tridva})-(\ref{eq: tritri})
for the massless matter fields
only in the presence of the Coulomb interaction
and $b$ treated as an external field.
Here we want to give the same solution but without
the additional assumption $A_{0}(0,t)=0$ used earlier.
For this reason the expressions for the eigenfunctions
and the spectrums given below are slightly different
from the ones in \cite{barsa94}.

The solution is
\ren
\begin{equation}
{\eta}_{1,n}^{c} =  \frac{1}{2{\rm L}}  \exp\{ -\frac{i}{2} e_1e_2
I_1(x_{-},x_{+}) - \frac{i}{2} (E_{1,n}^c
+ \frac{1}{2} (e_1+e_2) b ) x_{+} \},
\label{eq: tripet}
\end{equation}
\begin{equation}
{\eta}_{2,n}^c   =  \frac{1}{2{\rm L}}  \exp\{ -\frac{i}{2} e_1e_2
I_2(x_{-},x_{+}) - \frac{i}{2} (E_{2,n}^c
+ \frac{1}{2} (e_1-e_2) b ) x_{-} \},
\label{eq: trishest}
\end{equation}
where
\begin{eqnarray*}
I_1(x_{-},x_{+}) & \equiv & \frac{1}{2} x_{+} D_{+}(x_{+}|{\rm L}) +
x_{+} D_{-}(x_{-}|{\rm L}) - \frac{1}{24{\rm L}} x_{+}^3, \\
I_2(x_{-},x_{+}) & \equiv & \frac{1}{2} x_{-} D_{-}(x_{-}|{\rm L}) +
x_{-} D_{+}(x_{+}|{\rm L}) + \frac{1}{24{\rm L}} x_{-}^3 . \\
\end{eqnarray*}
The eigenvalues $E_{1,n}^c$, $E_{2,n}^c$ are determined by the
boundary conditions. From the boundary condition connecting
the values of ${\eta}_1$ at the points $(x_{-}=0$ , $x_{+}=0)$
and $(x_{-}=0$ , $x_{+}=2{\rm L})$ we get
\[
E_{1,n}^c = - \frac{1}{2{\rm L}} \int_0^{2{\rm L}} dz V(0,z)
+ \frac{2{\pi}n}{{\rm L}} - \frac{1}{2} (e_1+e_2)b,
\hspace{5 mm} n \in \cal Z,
\]
while the boundary conditions connecting the values of ${\eta}_2$
at $(x_{-}=0$ , $x_{+}=0)$ and $(x_{-} = \pm {\rm L}$,
$x_{+}={\rm L})$ give
\[
E_{2,n}^c = - \frac{1}{2{\rm L}} \int_{-{\rm L}}^{{\rm L}}
dz V(z,{\rm L}) + \frac{2{\pi}n}{{\rm L}} - \frac{1}{2} (e_1-e_2)b,
\hspace{5 mm} n \in \cal Z.
\]

For $V={\phi}$ (with the assumption $A_0(0,t)=0$ both parts
of the Coulomb potential ${\phi}_{+}$ and ${\phi}_{-}$ would
be asymmetric \cite{barsa94} ), we easily evaluate the
integrals, so the spectrums become
\[
E_{1,n}^c = - \frac{1}{3} e_1e_2 {\rm L} + \frac{2{\pi}}{\rm L} n
- \frac{1}{2} (e_1+e_2)b,
\]
\[
E_{2,n}^c = - \frac{5}{12} e_1e_2 {\rm L} + \frac{2{\pi}}{\rm L} n
- \frac{1}{2} (e_1-e_2)b.
\]
The eigenfunctions
${\eta}_{3,n}^c$ and ${\eta}_{4,n}^c$
are obtained from Eqs.(\ref{eq: tripet})-(\ref{eq: trishest})
 by making use of the
relations (3.4a-b), the corresponding spectrums being
\begin{eqnarray*}
E_{3,n}^c & = & -\frac{5}{12}e_1e_2{\rm L} + \frac{2\pi}{\rm L}n
+ \frac{1}{2}(e_1-e_2)b, \\
E_{4,n}^c & = & -\frac{1}{3}e_1e_2{\rm L} + \frac{2\pi}{\rm L}n
+ \frac{1}{2}(e_1+e_2)b, \hspace{5 mm} n \in \cal Z.
\end{eqnarray*}
The superscript "c"
indicates that the eigenfunctions
${\eta}_{i,n}^c$ and the eigenvalues $E_{i,n}^c$
represent the solution of our two-body problem in the
presence of the Coulomb interaction, but without the
self-potentials.

The boundary conditions fix also the phases
${\kappa}_1^{(i)}$ , ${\kappa}_2^{(i)}$ ,
\begin{eqnarray*}
{\kappa}_{1,n}^{(1)} = {\kappa}_{1,n}^{(4)}
= & - {\kappa}_{2,n}^{(1)} = & - {\kappa}_{2,n}^{(4)} = \frac{n}{2},\\
{\kappa}_{1,n}^{(2)} = {\kappa}_{1,n}^{(3)}
= & {\kappa}_{2,n}^{(2)} = & {\kappa}_{2,n}^{(3)} = \frac{n}{2}.\\
\end{eqnarray*}

\begin{center}

\subsection{The case ${\phi}_{(1)}^{\rm self} \neq 0$,
${\phi}_{(2)}^{\rm self} \neq 0$}

\end{center}

Let us now solve Eqs.(\ref{eq: tridva})-(\ref{eq: tritri})
 in the presence of the
self-potentials. In the self-field approach to
quantum electrodynamics in four dimensions the self-field
effects are calculated by an iteration procedure. To lowest
order of iteration we take the fields to be given by the
solutions without the self-energy terms, and the energies
to be shifted by a small amount:
\[
{\eta}_{i,n} = {\eta}_{i,n}^c,
\]
\[
E_n = E_n^c + \Delta E_n.
\]

But in two-dimensional quantum electrodynamics
for some problems we need not apply the iteration procedure,
because these problems can be solved exactly. We show below
that it is the case for the two-body problem in the
massless $2$-dim QED.

For the vanishing matter masses , the general structure
of the solutions to the Eqs.(\ref{eq: tridva})-(\ref{eq: tritri})
with the self-potentials is the same as of the corresponding solutions
in the case when the self-field effects are neglected.
All the solutions are exponents. So if these solutions
are normalized, they fulfil the conditions
\[
{\eta}_i^{\star} {\eta}_i = \frac{1}{4{\rm L}^2} ,
\hspace{5 mm} i=\overline{1,4}.
\]

The bilinear combinations
$\overline{\Phi} ({\gamma}^0 \otimes {\gamma}^0)
{\Phi}$ which enter the expressions for the self-potentials
can be therefore easily evaluated as
\[
\overline{\Phi} ({\gamma}^0 \otimes {\gamma}^0) {\Phi} =
\sum_{i=1}^4 {\eta}_i^{\star} {\eta}_i = \frac{1}{{\rm L}^2}.
\]

The self-potentials take the exact and closed form
\[
{\phi}_{(1)}^{\rm self} = \frac{e_1}{2{\rm L}} (x^2+\frac{{\rm L}^2}{2}),
\]
\[
{\phi}_{(2)}^{\rm self} = \frac{e_2}{2{\rm L}} (x^2+\frac{{\rm L}^2}{2}).
\]

In terms of the center of mass and relative coordinates
the self-field part of the total potential $V$
becomes
\[
e_1 {\phi}_{(1)}^{\rm self} + e_2 {\phi}_{(2)}^{\rm self} =
\frac{1}{8{\rm L}} (e_1^2 + e_2^2) (x_{+}^2 + x_{-}^2)
+ \frac{1}{4{\rm L}} (e_1^2 - e_2^2) x_{+}x_{-}
+ \frac{{\rm L}}{4} (e_1^2 + e_2^2),
\]
i.e., only for $e_1 = \pm e_2$ the self-potentials
do not destroy the additivity of the center of mass and
relative Hamiltonians.

With the self-potentials, Eqs.(\ref{eq: tridva})-
(\ref{eq: tritri})
are solved by the eigenfunctions
\[
{\eta}_{1,n} = \frac{1}{2{\rm L}}  \exp\{ - \frac{i}{2} e_1e_2
I_1(x_{-},x_{+}) - \frac{i}{2} (E_{1,n} + \frac{1}{2}
(e_1 + e_2)b) x_{+}
\]
\begin{equation}
- \frac{i}{2} e_1^2 J_1^{(1)}(x_{-},x_{+}) - \frac{i}{2}
e_2^2 J_1^{(2)}(x_{-},x_{+}) \} ,
\label{eq: trisem}
\end{equation}
\[
{\eta}_{2,n} = \frac{1}{2{\rm L}}  \exp\{ -\frac{i}{2} e_1e_2
I_2(x_{-},x_{+}|a) - \frac{i}{2} (E_{2,n} + \frac{1}{2}
(e_1 -e_2)b)(x_{-} - a{\rm L})
\]
\begin{equation}
- \frac{i}{2} e_1^2 J_2^{(1)}(x_{-},x_{+}|a) - \frac{i}{2}
e_2^2 J_2^{(2)}(x_{-},x_{+}|a) \},
\label{eq: trivosem}
\end{equation}
where
\[
I_2(x_{-},x_{+}|a) \equiv \frac{1}{2} x_{-}D_{-}(x_{-}|{\rm L})
+ (x_{-} - a{\rm L}) D_{+}(x_{+}|{\rm L}) + \frac{1}{24{\rm L}}
x_{-}^3 + I_2(0,0|a),
\]
and
$$
I_2(0,0|a) \equiv
\left\{ \begin{array}{cc}
\frac{{\rm L}^2}{24}a^2(2a+3) & {\rm for} \hspace{5 mm} a<0,\\
\vspace{2 mm}   \\
\frac{{\rm L}^2}{24}a^2(2a-9) & {\rm for} \hspace{5 mm} a>0,
\end{array}
\right.
$$
while
\[
J_1^{(1)}(x_{-},x_{+}) \equiv \frac{1}{24{\rm L}}
((x_{+}+x_{-})^3 - x_{-}^3) + \frac{\rm L}{4}x_{+},
\]
\[
J_1^{(2)}(x_{-},x_{+}) \equiv \frac{1}{24{\rm L}}
((x_{+}-x_{-})^3 + x_{-}^3) + \frac{\rm L}{4}x_{+},
\]
\[
J_2^{(1)}(x_{-},x_{+}|a) \equiv \frac{1}{24{\rm L}}
((x_{+} + x_{-})^3 - (x_{+} + a{\rm L})^3 )
+ \frac{\rm L}{4} (x_{-} - a{\rm L}),
\]
\[
J_2^{(2)}(x_{-},x_{+}|a) \equiv \frac{1}{24{\rm L}}
(-(x_{+} - x_{-})^3 + (x_{+} - a{\rm L})^3)
+ \frac{\rm L}{4} (x_{-} - a{\rm L}).
\]
The constant $a$  depends on the charges
$e_1,e_2$ , namely,
$a = (e_2-e_1)/(e_2+e_1)$  for $e_1 \neq \pm e_2$ and
$a=0$ for $e_1 = \pm e_2$.
In the Coulomb case when the self-potentials
are not taken into account, $a$ vanishes,and
\[
I_2(x_{-},x_{+}|0) =  I_2(x_{-},x_{+}),
\]
the functions ${\eta}_{1,n}$ , ${\eta}_{2,n}$ reducing
to the Coulomb eigenfunctions
${\eta}_{1,n}^c$ ,
${\eta}_{2,n}^c$.

The eigenvalues acquire a shift,
\new{a}
\begin{equation}
E_{1,n} = E_{1,n}^c + \Delta E_1,
\end{equation}
\add
\new{b}
\begin{equation}
E_{2,n} = E_{2,n}^c + \Delta E_2,
\end{equation}
which is nothing else than the self-energy
\[
\Delta E_1 \equiv - \frac{1}{2{\rm L}} \int_0^{2{\rm L}} dz
(e_1 {\phi}_{(1)}^{\rm self}(0,z) + e_2 {\phi}_{(2)}^{\rm self}(0,z)),
\]
\[
\Delta E_2 \equiv - \frac{1}{2{\rm L}} \int_{-{\rm L}}^{{\rm L}} dz
(e_1 {\phi}_{(1)}^{\rm self}(z,{\rm L}) + e_2 {\phi}_{(2)}^{\rm self}
(z,{\rm L})).
\]
The shift is the same for both spectrums
\[
\Delta E_1 = \Delta E_2 \equiv \Delta E =
- \frac{5}{12} e_1^2 {\rm L} -
\frac{5}{12} e_2^2 {\rm L}.
\]

The eigenfunctions ${\eta}_{3,n}$, ${\eta}_{4,n}$ are related to
${\eta}_{1,n}$, ${\eta}_{2,n}$ by Eqs.(3.4a-b),
the corresponding spectrums being shifted by the same
amount $\Delta E$.

The self-potentials contribute also to the boundary
conditions phases
\begin{eqnarray*}
{\kappa}_{1,n}^{(1)} = & - {\kappa}_{2,n}^{(1)}  & = \frac{n}{2}
+ {\kappa}_1^{\rm self}, \\
{\kappa}_{1,n}^{(2)} = & {\kappa}_{2,n}^{(2)} & = \frac{n}{2}
+ {\kappa}_2^{\rm self}, \\
{\kappa}_{1,n}^{(3)} = & {\kappa}_{2,n}^{(3)} & = \frac{n}{2}
- {\kappa}_2^{\rm self}, \\
{\kappa}_{1,n}^{(4)} = & - {\kappa}_{2,n}^{(4)} & = \frac{n}{2}
- {\kappa}_1^{\rm self},
\end{eqnarray*}
where
\begin{eqnarray*}
{\kappa}_1^{\rm self} & \equiv &
\frac{(e_2^2-e_1^2){\rm L}^2}{32{\pi}}, \\
{\kappa}_2^{\rm self} & \equiv & \frac{(e_1+e_2)^2{\rm L}^2}
{32{\pi}} a(2-a^2).
\end{eqnarray*}
The additional phases ${\kappa}_1^{\rm self}$,${\kappa}_2^{\rm self}$
vanish in the case $e_1= \pm e_2$.

With the Coulomb and self-interaction shifts, the spectrums
for $e_1=-e_2 \equiv e$ , for instance, become
\begin{eqnarray*}
E_{1,n} & = & -\frac{1}{2}e^2{\rm L} + \frac{2\pi}{\rm L}n,\\
E_{2,n} & = & -\frac{5}{12}e^2{\rm L} + \frac{2\pi}{\rm L}n -eb,
\hspace{5 mm} n \in \cal Z.\\
\end{eqnarray*}

Eqs.(3.7)-(3.9) represent the
complete and exact solution
of the two-body problem for the massless matter fields.

\renewcommand{\ren}{\renewcommand{\theequation}{4.\arabic{equation}}}
\renewcommand{\add}{\addtocounter{equation}{-1}}
\renewcommand{\new}[1]{\renewcommand{\theequation}{4.\arabic{equation}#1}}
\renewcommand{\set}{\setcounter{equation}{0}}

\begin{center}

\section{EQUIVALENCE}

\end{center}

\set

In the pair of Dirac equations formulation, the Coulomb
potential is included into the self-field terms.
With the assumptions that the composite matter field does
not depend on the relative time and $t_1=t_2$, the total
self-potential coincides with the corresponding one
in the single two-body equation formulation and can be
evaluated exactly as
\[
{\phi}^{\rm self}(x) = \frac{e_1+e_2}{2{\rm L}}
(x^2+\frac{{\rm L}^2}{2}).
\]
It is time-independent and satisfies the compatibility
condition (\ref{eq: dvaodindevet}).

The eigenvalue problem for the two-body Hamiltonian
reduces to the system of two equations for each component
of the composite field. For
${\eta}_1$, we have
\ren
\begin{equation}
({\Pi}^1 - V- E ){\eta}_1 =0,
\label{eq: cetodin}
\end{equation}
\begin{equation}
({\pi}^1 - U){\eta}_1 =0,
\label{eq: cetdva}
\end{equation}
where the last equation means the vanishing of the relative
energy, and
\[
V(x_{-},x_{+}) = e_1{\phi}^{\rm self}(\frac{x_{+}+x_{-}}{2}) +
e_2{\phi}^{\rm self}(\frac{x_{+}-x_{-}}{2}),
\]
\[
U(x_{-},x_{+}) \equiv e_1{\phi}^{\rm self}(\frac{x_{+}+x_{-}}{2}) -
e_2{\phi}^{\rm self}(\frac{x_{+}-x_{-}}{2}).
\]
The potentials $V$ and $U$ fulfil the relations
\new{a}
\begin{equation}
\frac{{\partial}V}{{\partial}x_{-}}=
\frac{{\partial}U}{{\partial}x_{+}},
\end{equation}
\add
\new{b}
\begin{equation}
\frac{{\partial}V}{{\partial}x_{+}}=
\frac{{\partial}U}{{\partial}x_{-}},
\end{equation}
the first one being the compatibility condition for
Eqs.(4.1) and (4.2).

The general solution of (4.1) is
\ren
\begin{equation}
{\eta}_1(x_{-},x_{+}) = {\chi}_1(x_{-}) \exp\{ - \frac{i}{2}
\int_{0}^{x_{+}} dz V(x_{-},z) - \frac{i}{2} (E+\frac{1}{2}
(e_1+e_2)b) x_{+} \},
\label{eq: cetcet}
\end{equation}
where ${\chi}_1$ is a function depending only on
the relative coordinate $x_{-}$.

Substituting this solution into (4.2) and using the relations
(4.3a-b), we get the equation for ${\chi}_1$,
\[
({\pi}^1 - U(x_{-},0) ) {\chi}_1 = 0,
\]
which is solved by
\ren
\begin{equation}
{\chi}_1(x_{-}) = \exp\{ - \frac{i}{2} \int_{0}^{x_{-}} dz
U(z,0) - \frac{1}{4} (e_1-e_2) b x_{-} \}.
\label{eq: cetpet}
\end{equation}
Although the solution (4.4)-(4.5) includes both potentials
$V$ and $U$ , only the potential $V$ contributes to the
eigenvalue spectrum. Indeed, ${\chi}_1(0)=1$
at the boundary points $(x_{-}=0,x_{+}=0)$
and $(x_{-}=0,x_{+}=2{\rm L})$. Since just the boundary
condition connecting the values of ${\eta}_1$
at these points determines the spectrum,
the potential $U$ drops out of this boundary condition,
and for the spectrum we get the same expression as in the
single two-body equation formulation.

For the second component ${\eta}_2$, the system of equations
for the eigenfunctions is
\new{a}
\begin{equation}
( {\pi}^1 + V - E) {\eta}_2 =0,
\end{equation}
\add
\new{b}
\begin{equation}
( {\Pi}^1 + U ) {\eta}_2 =0.
\end{equation}
The solution is given by
\ren
\begin{equation}
{\eta}_2(x_{-},x_{+}) = {\chi}_2(x_{+}) \exp \{ \frac{i}{2}
\int_{a{\rm L}}^{x_{-}} dz V(z,x_{+}) -
\frac{i}{2}
(E + \frac{1}{2}(e_1-e_2)b)(x_{-} -a{\rm L}) \},
\label{eq: cetsem}
\end{equation}
where
\begin{equation}
{\chi}_2(x_{+}) = \exp \{ \frac{i}{2} \int_0^{x_{+}} dz U(0,z)
- \frac{i}{4} (e_1+e_2)b x_{+} \}.
\label{eq: cetvosem}
\end{equation}

From the boundary conditions relating the points
$(x_{-}=0$, $x_{+}=0)$ and $(x_{-}= \pm {\rm L}$,
$x_{+}= {\rm L})$ and with ${\kappa}_1^{(2)} =
{\kappa}_2^{(2)}$ we get the following condition
determining the spectrum
\[
{\eta}_2(-{\rm L}|{\rm L}) = {\eta}_2 ({\rm L}|{\rm L}).
\]
In both parts of this condition we have the function
${\chi}_2(x_{+})$ taken at the same center of mass coordinate
$x_{+}={\rm L}$, so the potential $U$ drops out again.

Thus, in both formulations the
eigenvalue spectrums coincide. This
proves that the two formulations
are equivalent to each other in the spectrum problem.

\begin{center}

\section{DISCUSSION}

\end{center}

1. For $(1+1)$-dimensional self-field QED, we have presented
two different formulations of the two-body problem in
accordance with two different types of variational principles.
These two formulations are closely related but not identical.
In the first formulation we vary the action with respect to
the composite matter field and get a single two-body equation.
In the second formulation we require the action be stationary
with respect to the individual fields. This condition is stronger
and leads to a pair of equations for the composite field.
In addition to the two-body equation we have an equation
including the relative energy. While the single two-body
equation formulation is one-time formulation without any
restrictions on the self-potentials, the second formulation
has in general two time coordinates. Only for a special time
dependence of the self-potentials one of the time coordinates,
i.e., the relative time can be eliminated.

We have shown that for the massless matter fields the eigenvalue
spectrums of the two-body Hamiltonian in both formulations
coincide. The two formulations are therefore equivalent in the
spectrum problem. Nevertheless, the second formulation with two
compatible equations on the same composite field provides more
complete information about the eigenfunctions. The single
two-body equation does not fix the eigenfunctions uniquely.
We can multiply the components ${\eta}_1$ and ${\eta}_4$ by
an arbitrary function depending only on the relative
coordinate $x_{-}$ and the components ${\eta}_2$ and ${\eta}_3$
by one depending only on $x_{+}$.

2. We have proved that the relativistic two-body problem in
the massless two-dimensional quantum electrodynamics
is exactly soluble. In the single two-body equation
formulation, we have
solved the covariant two-body equation with both mutual
and self-interactions and found the eigenfunctions and the
spectrum of the two-body Hamiltonian.

For the massive matter fields, the eigenvalue problem for
the two-body Hamiltonian becomes essentially more
complicated. In this case, $\eta$'s are not decoupled
in the system of equations (3.2)-(3.3).
If we try to decouple them, then we arrive at a set of
second-order differential equations which can be solved
only in some approximation. We can take
the masses $m_1$ , $m_2$ as small parameters and consider
the mass contribution to the two-body Hamiltonian
eigenfunctions and eigenvalues as small corrections to
the corresponding eigenfunctions and eigenvalues for
the vanishing masses. The discussion of the massive
case will be given elsewhere.

There is an essential difference in the Coulomb and self-interaction
shifts in the spectrums. The Coulomb interaction shifts the
discrete energy spectrums by a value which is different for
$E_{1,n}$ and $E_{2,n}$ ($E_{4.n}$ and $E_{3,n}$) , the
difference being equal to $\frac{1}{12}e_1e_2{\rm L}$ , while
the self-interaction shift is the same for all four spectrums.

If we call formally the first and second components of the
two-component fields ${\psi}_k$ as "up" and "down" components,
then the two-body system states described by ${\eta}_{i,n}$
$(i=\overline{1,4})$ can be interpreted correspondingly as
"up-up", "up-down", "down-up" and "down-down" states. The
Coulomb interaction shift is therefore the same for the
"up-up" and "down-down" states and takes a different value
for the "up-down" and "down-up" states. Thus we can recognize
the effects of spin-spin interactions in the first
quantized theory.

For arbitrary values of $e_1 , e_2$, all the spectrums $E_{i,n}$
depend on the global electromagnetic field degree
of freedom $b$. The global degree of freedom contribution
to the spectrums is specific to models defined on the circle.
For models on the line, the electromagnetic field has
neither local nor global physical degrees of freedom
and so can be eliminated completely from the two-body
Hamiltonian.

For $e_1=-e_2$, only the spectrums $E_{2,n}$ and $E_{3,n}$
corresponding to the "up-down" and "down-up" states ,
and for $e_1=e_2$ only $E_{1,n}$ and $E_{4,n}$
corresponding to the "up-up" and "down-down" states
depend on $b$.

3. The standard SM with a single matter field of charge $e$ is
equivalent to the theory of a free scalar field with mass
$e^2/{\pi}$ \cite{js62}. In our work, we have looked at the
SM from a different point of view. We have constructed the
mass spectrum for the model with two matter fields. The
spectrum obtained does not contain the boson of the SM.
This result is not surprising. It is well known from the
second quantized version of $(1+1)$-dimensional QED that
only on light front the SM boson can be represented as
a bound state of two fields, fermion and antifermion
\cite{berg77,brod98}. The study of the self-field SM on
light front was given in \cite{fms98}.

In $(3+1)$-dimensions the formulations presented above
give us a possibility to make calculations for real two-body
systems. In the self-field approach, the single two-body
equation formulation was used in \cite{oba86} to calculate
the energy spectrum for positronium and muonium. In the
framework of the constraint approach, the pair of Dirac
equations formulation was applied to the phenomenological
calculation of the $q-\overline{q}$ meson bound state
spectrum as well as to the study of the dynamics of
quarkonium systems \cite{cra84,saz86}. Both formulations
produce results which agree with experiment. However, to
clarify the difference between the two formulations in
$(3+1)$-dimensions an analytical work along the lines
given in the present paper for $(1+1)$-dimensions is
needed.

\end{document}